\documentclass[review]{elsarticle}

\usepackage{amsmath}
\usepackage{graphicx}
\usepackage{wrapfig}
\usepackage{todonotes}

\graphicspath{{img/}}

\usepackage{hyperref}

\journal{Journal of Computational Science}

\begin{document}

\begin{frontmatter}

\title{Flow through time--evolving porous media: swelling and erosion}

\author[cftc,ulisboa]{Andr\'{e} F.~V. Matias\corref{correspondingauthor}}
\cortext[correspondingauthor]{Corresponding author, +351~21~750~0989, Office 8.6.14, Faculdade de Ci\^{e}ncias, Universidade de Lisboa, 1749--016 Lisboa, Portugal}
\ead{afmatias@fc.ul.pt}
\author[cftc,ulisboa]{Rodrigo C.~V. Coelho}
\ead{rcvcoelho@fc.ul.pt}
\author[ceara]{Jos\'{e} S. Andrade Jr.}
\ead{soares@fisica.ufc.br}
\author[cftc,ulisboa]{Nuno A.~M. Ara\'{u}jo}
\ead{nmaraujo@fc.ul.pt}

\address[cftc]{Centro de F\'{i}sica Te\'{o}rica e Computacional, Faculdade de Ci\^{e}ncias, Universidade de Lisboa, 1749--016 Lisboa, Portugal}
\address[ulisboa]{Departamento de F\'{i}sica, Faculdade de Ci\^{e}ncias, Universidade de Lisboa, 1749--016 Lisboa, Portugal}
\address[ceara]{Departamento de F\'{i}sica, Universidade Federal do Cear\'{a}, 60451--970, Fortaleza, Cear\'{a}, Brazil}

\begin{abstract}

The flow through a porous medium strongly depends on the boundary conditions, very often assumed to be static. Here, we consider changes in the medium due to swelling and erosion and extend existing lattice--Boltzmann models to include both. We study two boundary conditions: a constant pressure drop and a constant flow rate. For a constant flow rate, the steady state depends solely on the erosion dynamics while for a constant pressure drop it depends also on the timescale of swelling. We analyze the competition between swelling and erosion and identify a transition between regimes where either swelling or erosion dominate.

\end{abstract}

\begin{keyword}
Lattice--Boltzmann method \sep porous media \sep swelling \sep erosion.
\end{keyword}

\end{frontmatter}

\section{Introduction}

In many natural and industrial processes such as groundwater transport, food grain drying, and oil or coffee extraction, a fluid flows through a network of channels (porous medium) that significantly constrain the flow~\cite{DeBoer2000, Sahimi1993, Berkowitz2002, Coelho2016}. Several types of fluid/medium interaction are possible. For example, in chemical filtration, the solid matrix interacts selectively with the dispersed chemical compounds retaining some of them~\cite{Bear2018, OMelia1967, Faust2018}. By contrast, in coffee extraction or in aquifers molecules are extracted from the medium and transported by the fluid flow~\cite{Ellero2019, Abriola1985, Berkowitz2002}. In the latter, besides extraction, the particles of the medium also absorb the fluid and swell, significantly reducing the space for the flow~\cite{Aksu2015, Bakhshian2017, Mateus2007}. Erosion of the solid matrix occurs from the liquid--solid interaction, as for example in coffee extraction, leading to the transport of fines downstream to the cup~\cite{Illy2005} or in oil extraction, where the shear stress causes erosion of the well leading to the transport of sand and reducing the yield of the extraction~\cite{Ghassemi2015}. All these changes are expected to affect the flow~\cite{Jager2017}.

Previous research works have focused on the impact of erosion and deposition. It was shown, experimentally~\cite{Mahadevan2012, Kudrolli2016} and numerically~\cite{Jager2017, Derr2020}, that erosion and deposition can lead to channelization of porous media. These changes in the medium can even lead to clogging of pipes~\cite{Sendekie2016, Jager2018, Filho2016}. The flow of groundwater can cause erosion and fluidization of the soil leading to landslides~\cite{Lobkovsky2005}. All these studies assume that erosion and deposition depend linearly on the shear stress~\cite{Bonelli2006}. In the numerical works, special attention is given to the location of the solid boundaries, given the sensitivity to this detail. Thus, several methods have been employed to improve the boundary location estimation such as using interpolated boundary conditions~\cite{Mei2002} or using more sophisticated collision operators~\cite{Ginzburg2008}.

The swelling of the solid matrix is mainly studied experimentally in food science~\cite{Mateus2007} and fluid flow in aquifers~\cite{Bakhshian2017} and a theoretical understanding of the process is still elusive. Different from erosion, swelling is a process regulated by the diffusion of liquid into the solid matrix and so it does not depend significantly on the fluid flow.

Here, we explore the competition between swelling and erosion in porous media performing simulations with the lattice--Boltzmann method. This competition differs from the above mentioned studies on erosion--deposition competition \cite{Jager2017, Derr2020, Sendekie2016, Jager2018} since, while erosion changes the local properties of the solid matrix, the swelling directly affects it globally through volume and surface area changes. We assume that erosion depends linearly on the wall shear stress and that there is an erosion threshold that prevents erosion for low shear stress, as implemented on previous works~\cite{Jager2017}. We extend the erosion implementation to include swelling. The fluid is set into motion by imposing two types of boundary conditions: a constant pressure drop or a constant flow rate. When a constant flow rate is imposed, the dynamics is solely determined by erosion, while for a constant pressure drop there is a clear competition between swelling and erosion, which depends on the timescale of swelling. For a constant pressure drop, we determine how the total mass eroded depends on the erosion and swelling parameters. Depending on the model parameters, we find two regimes: one where the medium is drastically eroded and one where no erosion occurs.

The paper is organized as follows. We start by describing the model and numerical simulations in Sec.~\ref{sec:model}, namely the details about the lattice--Boltzmann method, used to simulate the fluid flow and the swelling and erosion implementation. In Sec.~\ref{sec:validation} we explore certain limits to validate the model. In Sec.~\ref{sec:results} we study the competition between swelling and erosion for the two boundary conditions. We draw some conclusions in Sec.~\ref{sec:conclusion}.

\section{Model description}
\label{sec:model}

\subsection{Lattice--Boltzmann Method}

The numerical simulations of the fluid were performed using the lattice--Boltzmann method (LBM), where the Boltzmann equation is discretized and solved numerically. We use the D3Q19 discretization, where space is discretized into a regular cubic grid in 3D, of linear lengths $L_x$, $L_y$ and $L_z$, and velocity is discretized into a set of 19 velocity vectors~\cite{Kruger2017, Succi2018}. It is well documented on the literature that this lattice choice is suitable to simulate laminar flows~\cite{Kruger2017}. The discretized Boltzmann equation is
\begin{equation}
	\frac{f_i(\vec{x}+\vec{c}_i \Delta t, t + \Delta t) - f_i(\vec{x}, t)}{\Delta t} = \left(\frac{\partial f}{\partial t}\right)_\text{coll}\text{ ,}
	\label{eq:boltzmann_eq}
\end{equation}
where $f_i(\vec{x}, t)$ is the distribution function in direction $\vec{c}_i$ at the node with position $\vec{x}$ and time $t$, $\Delta t$ is the time interval between iterations, and $\vec{c}_i$ are the discretized velocities. For the collision term, we consider the two--relaxation--time operator (TRT),
\begin{equation}
	\left(\frac{\partial f}{\partial t}\right)_\text{coll} = -\frac{f_i^+ - f^{\text{eq}+}_i}{\tau^+} - \frac{f_i^- - f^{\text{eq}-}_i}{\tau^-} \text{ ,}
\end{equation}
where the $+$ and $-$ upper scripts indicate the symmetric and anti--symmetric part of the distribution function:
\begin{align}
	f_i^+ &= \frac{f_i+f_{\bar{i}}}{2} \text{ ,} &f_i^- =& \frac{f_i-f_{\bar{i}}}{2} \text{ ,}
	\\
	f_i^{\text{eq}+} &= \frac{f_i^\text{eq}+f_{\bar{i}}^\text{eq}}{2} \text{ ,} &f_i^{\text{eq}-} =& \frac{f_i^\text{eq}-f_{\bar{i}}^\text{eq}}{2} \text{ ,}
	\label{eq:symm_antisym}
\end{align}
where $f_{\bar{i}}$ is the distribution on the direction $\vec{c}_{\bar{i}}=-\vec{c}_i$, $\tau^+$ is related to the kinematic viscosity of the fluid $\nu = c_s^2\left(\tau^+ - \Delta t/2\right)$, and $c_s=(1/\sqrt{3})(\Delta x / \Delta t)$ is the sound speed in the discretized lattice. Our results are expressed in lattice units: the space between lattice sites $\Delta x$ and time step $\Delta t$ are both unitary. Throughout the work we set $\tau^+=0.8$, which corresponds to $\nu=0.1$, well within the regime of laminar flow \cite{Faber1995, Acheson1990, Landau2013}. The second relaxation time $\tau^-$ is related to $\tau^+$ as
\begin{equation}
	\Lambda = \left(\tau^+ - \frac{1}{2}\right)\left(\tau^- - \frac{1}{2}\right) \text{ ,}
\end{equation}
where $\Lambda$ is a free parameter. By using the TRT operator and choosing $\Lambda=3/16$, we ensure that the effective position of the boundary is independent of the fluid viscosity~\cite{Ginzburg2008, Kruger2017}. The equilibrium distribution $f^\text{eq}_i$ is
\begin{equation}
	f^\text{eq}_i = w_i \rho \left(1+\frac{\vec{c}_i\cdot\vec{u}}{c^2_s} - \frac{\vec{u} \cdot \vec{u}}{2c^2_s} + \frac{(\vec{c}_i\cdot\vec{u})^2}{2c^4_s}\right) \text{ ,}
	\label{eq:MBdist}
\end{equation}
where $w_i$ are weights related to the discretized velocities that are required to ensure mass conservation. The weights and velocity vectors follow the standard implementation of the D3Q19 lattice~\cite{Kruger2017}. The macroscopic variables of the fluid, density $\rho$ and velocity $\vec{u}$ are determined from the distribution functions with $\rho = \sum_i f_i$ and $\rho \vec{u} = \sum_i f_i \vec{c}_i$.

The solid matrix is defined by nodes that cover the solid region. Each node has a mixture of fluid and solid measured by the solid fraction. Depending on the solid fraction of each node, there are three types of nodes: nodes that contain only fluid, nodes that contain only solid, and interface nodes that contain both solid and fluid. This translates into a scalar field
\begin{equation}
	s(\vec{x})=
	\begin{cases}
		0     & \quad \text{fluid node,}\\
		]0,1] & \quad \text{interface node,}\\
		1     & \quad \text{solid node.}
	\end{cases}
\end{equation}
If for a node $s(\vec{x})=1$, then the node is considered interface or solid depending if it has a fluid neighbor or not, respectively. On the interface nodes, no--slip boundary conditions are imposed using Mei's ghost nodes method, which allows continuum changes in the boundary position~\cite{Mei2002}. This method uses the distribution function of the interface node and its neighbors to extrapolate the distribution function coming out of the boundary.

At the inlet and outlet, the distributions are controlled such that the desired boundary condition is imposed. For a constant pressure drop $\Delta p$ across $L_y$, the density of the nodes is fixed such that $\Delta p = c_s^2 (\rho_\text{out} - \rho_\text{in})$. Thus, we assume that the fluid is slightly compressible, regardless the variation of density do not have a meaningful impact on the results for the small dimensions we employ, $\approx300$ nodes in length. For a constant flow rate $Q_\text{fix}$ on the inlet and outlet, with area $L_x\times L_y$, the velocity of the nodes is fixed such that $Q_\text{fix} = v_\text{fix} \times L_x \times L_z$. To fix the density (or velocity) at the boundaries, the distributions at the inlet are determined based on the imposed density (or velocity) and on the known distribution functions of the node, as proposed by Kutay \textit{et al.}~\cite{Kutay2006}.

\subsection{Surface area}
\label{sec:normal_surface_area}

\begin{figure}
	\centering
	\includegraphics[width=0.3\textwidth]{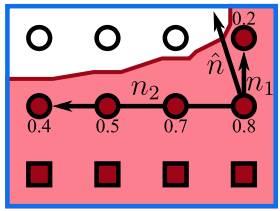}
	\caption{Schematic of the node surface area calculation. The white circles are fluid nodes, the red ones are interface nodes, and the red squares are solid nodes. The numbers next to the red circles represent the volume fraction $s(\vec{x})$. The diagonal vector is the normal vector $\hat{n}$ and the two orthogonal vectors are the rescaled components of the normal, $n_1$ and $n_2$.}
	\label{fig:surfacearea}
\end{figure}

Assuming that the boundary between fluid and solid is composed of straight lines that are at a distance $s(\vec{x})$ from the interface node, as is represented in Fig.~\ref{fig:surfacearea}, we determine the normal to the surface of a particle with the normalized weighted average of the neighboring fraction
\begin{equation}
	\hat{n}(\vec{x})=\frac{\sum_i s(\vec{x}+\vec{c}_i)\vec{c}_i}{\lVert\sum_i s(\vec{x}+\vec{c}_i)\vec{c}_i\rVert} \text{ ,}
	\label{eq:normal}
\end{equation}
where the sum is over all velocities $\vec{c}_i$.

The surface area of each node is obtained based on the orientation of the vector normal to the surface, see Eq.~\eqref{eq:normal}. We rescale the vector such that the smallest non--zero component, $n_1$, has unitary length, as seen in Fig.~\ref{fig:surfacearea}. With the three rescaled components, $n_1=1$, $n_2$, and $n_3$, where $n_1\leq n_2 \leq n_3$, we determine the area of a quadrilateral that is on top of the interface node and parallel to the boundary. This quadrilateral, of area $\sqrt{1+n_2^2}\sqrt{1+n_3^2}$, covers $(1+n_2)(1+n_3)$ interface nodes. The surface area per node is
\begin{equation}
	S_i = \frac{\sqrt{1+n_2^2}}{(1+n_2)} \frac{\sqrt{1+n_3^2}}{(1+n_3)} \text{ .}
	\label{eq:surface_area}
\end{equation}
Figure~\ref{fig:surfacearea} shows a 2D example of the surface area calculation. In this example, $n_2=3$ and $n_3=0$ and so a line with length $\sqrt{1+n_2^2}=\sqrt{10}$ spans $1+n_2=4$ interface nodes (represented as red circles). Notice that the two edge nodes only count as half node since they are shared with the next interface node.

\subsection{Swelling}

Mateus \textit{et al.}~\cite{Mateus2007} have shown that the volume increase of spherical coffee beans due to water absorption is well described by the solution of the advection--diffusion equation for an inflow of liquid into a spherical domain~\cite{Crank1979}
\begin{equation}
	\frac{V_t-V_0}{V_\infty-V_0}=1-\frac{6}{\pi^2}\sum_{n=1}^\infty \frac{1}{n^2}\exp \left(-\frac{Dn^2\pi^2t}{R^2}\right) \text{ ,}
	\label{eq:swell_teo}
\end{equation}
where $V_0$, $V_t$, and $V_\infty$ are the volume of the sphere at initial time, at time $t$, and asymptotically, respectively, $D$ is the diffusion coefficient of water inside the particles, and $R$ is the initial radius of the sphere. To describe swelling, based on this experimental result~\cite{Mateus2007}, we consider
\begin{equation}
	\frac{V(t)}{V_0}= 1 + (\alpha - 1) \left(1 - e^{-t/\tau_s}\right) \text{ ,}
	\label{eq:swell_simple}
\end{equation}
where $\alpha=2$ adjusts to the reported results, $\alpha V_0$ sets the final volume of the particles and $\tau_s$ sets the timescale of swelling. To obtain the volume change on each node, we assume that at each time step the volume of the particle swells by a thickness $h$. Thus
\begin{equation}
	\dot{h} = \frac{V_0}{S} \frac{\alpha - 1 }{\tau_s} e^{-t/\tau_s} \text{ ,}
\end{equation}
where $S$ is the total surface area of the particle. The increase in volume on each interface node is determined by multiplying the absorbed volume by the surface area of each interface node, $\dot{s} = S_i \dot{h}$ (see the definition of node surface area in Sec.~\ref{sec:normal_surface_area}). With this approach, the volume gained due to liquid absorption is composed mainly of liquid, and so the solid mass present in this volume is negligible when compared to the amount of liquid. This approximation holds as long as $V_0$ remains constant. With the increase in node volume some interface nodes became solid. When this happens, the distributions $f_i$ of the node are set to zero and its fluid neighbors become interface nodes. The transition to $f_i=0$ is abrupt. We observe that the fluid density remains constant and the velocity vanishes as the node volume increases.

\subsection{Erosion}

The model for erosion is based on the empirical observation~\cite{Parker2000} that the erosion rate is proportional to the wall shear stress
\begin{equation}
	T_w=\mu \frac{\partial u}{\partial \hat{n}}\bigg|_\text{surface} \text{ ,}
	\label{eq:WSS}
\end{equation}
where $\mu$ is the shear viscosity ($\mu = \nu \rho$), $u$ is the velocity parallel to the surface, and $\hat{n}$ is the surface unit vector. The experimental report also suggest a threshold in the shear stress for erosion $T_\text{er}$~\cite{Parker2000}. Thus, the amount of mass removed per unit area is given by~\cite{Jager2017}
\begin{equation}
	\dot{m} = 
	\begin{cases}
		-\kappa_\text{er} (T_w - T_\text{er}) & \quad \text{if } T_w  >  T_\text{er}\\
		0                                     & \quad \text{if } T_w \le T_\text{er}\\	
	\end{cases}
	\text{ ,}
	\label{eq:erosion}
\end{equation}
where $\kappa_\text{er}$ is the erosion rate. To numerically calculate the shear stress, we use the deviatoric shear stress tensor
\begin{equation}
	\sigma_{ab} = \mu \left(\frac{\partial u_a}{\partial x_b} + \frac{\partial u_b}{\partial x_a}\right) = \left(1 - \frac{\Delta t}{2\tau^+}\right) \sum_i (f_i-f_i^\text{eq})(\vec{c}_i)_a(\vec{c}_i)_b \text{ ,}
\end{equation}
where $(\vec{c}_i)_a$ is the $a$ component of $\vec{c}_i$, and the sum is over all velocities $\vec{c}_i$. Hence the shear stress is $\vec{T} = \hat{n} \cdot \sigma$. The shear stress parallel to the surface is $T_w = \sqrt{\vec{T}^2 - \left(\vec{T} \cdot \hat{n} \right)^2}$. The surface area of the interface node $S_i$ (see the definition of node surface area in Sec.~\ref{sec:normal_surface_area}) determines the volume change of the node $\dot{s} = S_i \dot{m} / \rho_s$, where $\rho_s$ is the density of the solid matrix. For simplicity we consider that the solid matrix is mainly composed of liquid, this is the case for the erosion that occurs during an espresso extraction where the water flow erodes organic matter, mainly composed of water. Thus, the density of the removed mass equals the fluid density, which we set to unity, and we do not track the eroded mass once it is dragged by the flow~\cite{PereiraNunes2016}.

The classification of a node changes over time. When a solid node becomes interface its fluid density is set to be the average density of the neighbors with fluid, its velocity to zero, the equilibrium distributions are determined with its density and velocity and the non--equilibrium part of the distribution functions are the same as the ones of the closest neighbor to a line in the direction of $\hat{n}$. To ensure that there is always a solid boundary, we assume that the inner core of the obstacles cannot be eroded.

\section{Numerical validation of the model}
\label{sec:validation}

\subsection{Swelling}

\begin{figure}
	\centering
	\includegraphics[width=0.5\textwidth]{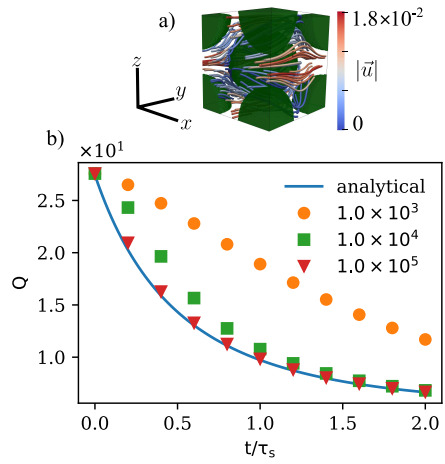}
	\caption{Flow past a BCC lattice of swelling spheres with a constant pressure drop. a) Simulation snapshot with the solid spheres in green and with the velocity streamlines colored with the velocity magnitude. There are $128^3$ nodes, the initial sphere radius is 40 nodes and a constant pressure drop of $\Delta p = 10^{-6}L_y$ is imposed on the $y$ direction. b) Flow rate $Q$ as a function of time normalized by the timescale of swelling. The different colors and markers correspond to different timescales of swelling. The analytical result, assuming steady state flow for each particle volume Eq.~\ref{eq:darcy}, is plotted with a blue line.}
	\label{fig:swellingval}
\end{figure}

We first consider the flow across a BCC lattice of spheres of radius $R$ with period $L$, as represented in the inset of Fig.~\ref{fig:swellingval}. Due to the flow, the drag on each sphere is
\begin{equation}
	d^*=\sum_{s=0}^{30} q_s \chi^s \text{ ,}
\end{equation}
where $\chi=R/R_\text{max}$ is the ratio between the sphere radius and the radius when two spheres touch, and $q_s$ are coefficients that can be found in Ref.~\cite{Sangani1982}. With the drag on each sphere, we can determine the permeability of the medium from
\begin{equation}
	k = \frac{L^2}{6\pi (R/L)} \frac{1}{2d^*(\chi)} \text{ ,}
	\label{eq:perma_BCC}
\end{equation}
where the factor 2 comes from the fact that there are two spheres on the elementary unit of a BCC lattice~\cite{Pan2006}. To validate the numerical model, we compare the flow rate predicted by the Darcy's law
\begin{equation}
	Q = -L^2 \frac{k}{\mu} \frac{\Delta p}{L}
	\label{eq:darcy}
\end{equation}
where $\mu$ is the shear viscosity ($\mu = \nu \rho$), with the one obtained numerically. We consider a cubic domain with 128 nodes per side, a constant pressure drop of $\Delta p = 10^{-6}L_y$ across the $y$ direction, and a timescale of swelling $\tau_s$. The spheres start with $R=40$, as represented in Fig.~\ref{fig:swellingval}a). The volume of the particles increases in time, changing their radius according to
\begin{equation}
	R(t) = \left( \frac{3}{4\pi} V(t) \right)^{1/3} \text{ .}
\end{equation}
This changes the permeability of the medium which results in a lower flow rate, see blue line in Fig.~\ref{fig:swellingval}b).

Figure~\ref{fig:swellingval}b) shows the time evolution of the flow rate for different timescales of swelling. The $x$ axis is dimensionless so that we can compare different timescales of swelling. Initially, the flow rate is independent of the swelling timescale and in agreement with the value predicted analytically by the Darcy's law. As the particles swell the values obtained numerically deviate from the predicted ones, especially for low values of the timescale of swelling. For significantly large timescales, the flow is quasi--steady, which means that the changes in the solid are much slower than the time it takes for the fluid to adjust to such changes. As we decrease $\tau_s$ the boundary of the spheres evolves faster, and we observe that the fluid nodes do not converge as expected. As a result, the numerical permeability differs from the analytical one, Eq.~\eqref{eq:darcy}, since the later was obtained with the assumptions of a steady state flow for each particle volume. In our simulations we assume quasi--steady flows, but increasing the timescale of swelling also increases the computational cost. Thus, we need a compromise between flow convergence and computational effort, and so we considered $\tau_s\geq10^4$.

\subsection{Erosion}

\begin{figure}
	\centering
	\includegraphics[width=0.5\textwidth]{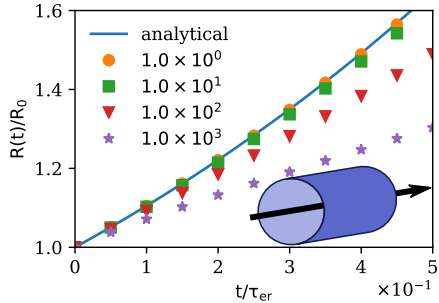}
	\caption{Pipe erosion with a constant pressure drop as a function of the dimensionless time. The different colors and markers correspond to different erosion rates. The erosion threshold is zero and the initial radius of the tube is $R_0=32$, schematized in the inset. The analytical result, Eq.~\eqref{eq:ero_teo}, is plotted with a blue line.}
	\label{fig:erosion_val}
\end{figure}

To validate the model for erosion, we consider the dynamics on a pipe. An analytic solution is possible using the Hagen--Poiseuille equation
\begin{equation}
	u(r)=-\frac{1}{{4\mu}}\frac{\Delta p}{L}(R^2-r^2) \text{ ,}
	\label{eq:poiseuille}
\end{equation}
where $\mu=\rho \nu$ is the shear viscosity, $R$ the tube radius, and $r$ the radial coordinate. Combining this equation with Eq.~\eqref{eq:WSS}, we obtain that the wall shear stress is
\begin{equation}
	T_w = \frac{\Delta p}{2L} R \text{ .}
	\label{eq:WSS_circles}
\end{equation}
With the solid density $\rho_s$ we can transform Eq.~\ref{eq:erosion} into the evolution of the pipe radius. Combining the radius evolution with the wall shear stress, in the absence of an erosion threshold ($T_\text{er}=0$), we obtain
\begin{equation}
	\frac{R(t)}{R_0}=e^{t/\tau_\text{er}} \text{ ,}
	\label{eq:ero_teo}
\end{equation}
where $R_0$ is the initial pipe radius and $\tau_\text{er}=2 (\rho_s / \kappa_\text{er}) (L / \Delta p)$ is the timescale of erosion. In Fig.~\ref{fig:erosion_val}, we compare this expression with the numerical results for $R_0=32$. The numerical results agree with the analytical expression, especially for small $\kappa_\text{er}$ and small $t/\tau_\text{er}$. As the radius of the pipe increases the shear stress increases which enhances erosion, thus the approximation of a quasi--steady flow worsens with time.

\section{Competition between swelling and erosion}
\label{sec:results}

Now we study the competition between swelling and erosion for two boundary conditions: a constant pressure drop and a constant flow rate. We fix the erosion rate $\kappa_\text{er}=1$, setting the timescale, and let the fluid reach the steady state before turning on swelling and erosion. This was achieved by simulating the fluid flow until the maximum relative changes in velocity, at each iteration, are smaller than $10^{-8}$. 

\subsection{Constant pressure drop}

To study the dynamics under a constant pressure drop, we fix $\Delta p = 10^{-5}L_y$ and vary the erosion threshold $T_\text{er}$. We first consider parallel plates, for which an analytical solution can be derived, and then a set of circular obstacles.

\subsubsection{Parallel plates}

\begin{figure}
	\centering
	\includegraphics[width=0.5\textwidth]{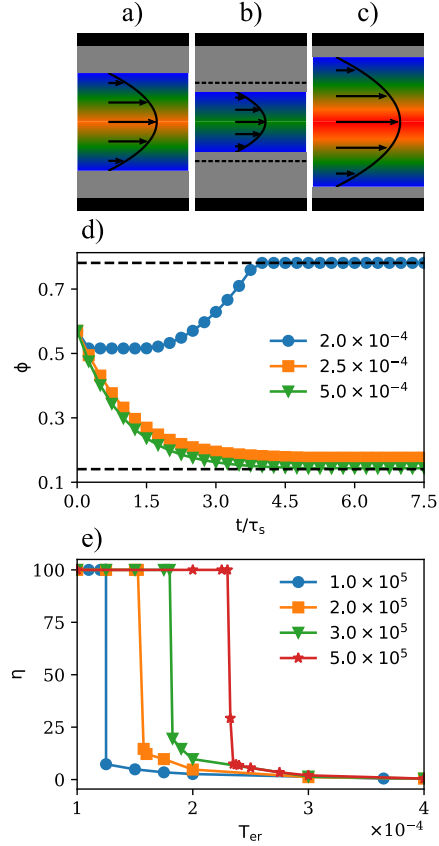}
	\caption{Erosion and swelling of two parallel plates with a constant pressure drop. a)-c) Contains three schematic examples of the shear stress. The colors and the arrows represent the velocity field (blue and red represent low and high velocity respectively). The parabola reflects the 2D version of the Hagen--Poiseuille equation, Eq.~\eqref{eq:poiseuille}. The dashed line in b) marks the minimum size gap required for the shear stress to overcome the erosion threshold; d) contains the evolution of the porosity $\phi$ for fixed timescale of swelling, $\tau_s = 5\times10^5$. The different colors and markers correspond to different erosion thresholds. e) contains the removal efficiency $\eta$ as a function of the erosion threshold. The different colors and markers correspond to different for different timescales of swelling. As we increase the timescale of swelling, the erosion threshold required for complete erosion ($\eta=100\%$) increases.}
	\label{fig:parallel}
\end{figure}

The parallel plates are made of two obstacles such that $s(\vec{x})=1$ for $x \le l_\text{min}$ and $x\ge l_\text{max}$. Thus, we have symmetry along the $y$ and $z$ direction, and so we set $L_y=L_z=1$ and impose periodic boundaries along the $y$ and $z$ directions. On the $x$ direction there are 128 nodes. The tortuosity is unitary meaning the result of a pressure drop is equivalent to a body force of magnitude $f_y=\Delta p / L_y$. We chose $f_y=10^{-5}$. We measured the porosity $\phi$, the fraction of the total volume that is not solid
\begin{equation}
	\phi = 1 - \frac{\sum_i^N s(\vec{x}_i)}{N} \text{ ,}
\end{equation}
where $N$ is the total number of nodes. Initially each plate has a volume of 26.5 nodes, thus initially the porosity is $\phi_i = 0.58$. Due to swelling the volume of the plates can double ($\alpha = 2$ in Eq.~\eqref{eq:swell_simple}), corresponding to the minimum porosity $\phi_\text{min}=0.17$. Due to erosion, the volume of the plates can decrease until it reaches a volume of 13 nodes, which sets the maximum porosity to $\phi_\text{max}=0.80$.

The wall shear stress on the parallel plates is similar to the case of a tube, Eq.~\eqref{eq:WSS_circles}. The difference is that the tube radius is replaced by the gap between the plates $l$, hence the shear stress is proportional to the imposed pressure drop and to the size of the gap between the plates $T_w \propto \Delta p \times l$. When the plates swell, the gap decreases and so it does the wall shear stress, which reduces the erosion. When the plates erode, the opposite happens. Thus, swelling creates a negative feedback on erosion, while erosion creates a positive feedback. Figures~\ref{fig:parallel}a)-c) are examples for three different values of the shear stress. Since the wall shear stress decreases with the gap between the plates, there is a size $l^*$, marked with a dashed line in Fig.~\ref{fig:parallel}b), such that the wall shear stress equals the erosion threshold. If the gap is smaller than this length, erosion stops.

In Fig.~\ref{fig:parallel}d) we plot the time evolution of porosity for different erosion thresholds. When the erosion threshold is large (green triangles), $l^*$ is larger than the initial gap size, thus no erosion occurs. Swelling dominates the dynamics and the porosity decreases until it reaches $\phi_\text{min}$. For small erosion threshold (blue circles), erosion dominates and the porosity increases, reaching $\phi_\text{max}$. The orange squares in Fig.~\ref{fig:parallel}d) correspond to $l^*$ slightly smaller than the initial gap. In this case, there is erosion at the beginning but swelling causes the gap size to evolve beyond $l^*$ at which point erosion stops. Thus, the dynamics evolves into a situation where the final porosity is between the maximum and the minimum value. In the cases where erosion can compensate swelling, such that $l^*$ is never crossed, everything is eroded, given the exponential decrease in swelling, Eq.~\eqref{eq:swell_simple}.

The porosity in the steady state was studied by plotting the removal efficiency $\eta$ which is the ratio between eroded volume and total volume that is possible to erode $s_\text{soft}$
\begin{equation}
	\eta = \left(1 - \frac{s_\text{final}-s_\text{liquid}-s_\text{min}}{s_\text{soft}}\right)\times 100\% \text{ ,}
	\label{eq:eta}
\end{equation}
where $s_\text{final}$, $s_\text{liquid}$, and $s_\text{min}$ are, respectively, the final volume of the solid matrix, the volume of the absorbed liquid and the total volume of the nodes that do not erode. The efficiency is zero when no volume is eroded and maximal when all possible volume is eroded. In Fig.~\ref{fig:parallel}e) the removal efficiency is plotted as a function of the erosion threshold for several timescales of swelling. For small erosion threshold, erosion dominates, and the maximum porosity is reached. On the other extreme swelling dominates and no erosion occurs. In between there is a decrease from maximum efficiency to zero. The value of $T_\text{er}$ at which the efficiency decreases depends on the timescale of swelling.

\begin{figure}
	\centering
	\includegraphics[width=0.7\linewidth]{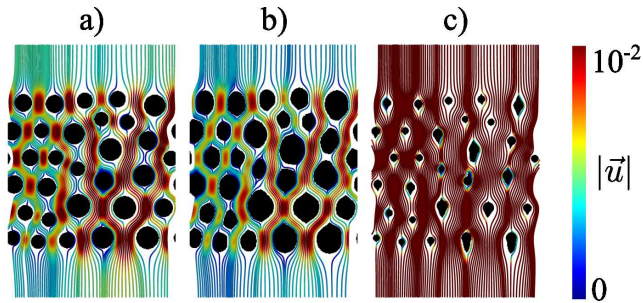}
	\caption{Three snapshots of a simulation with a constant pressure drop of a sample used in our analysis with velocity streamlines. Red lines correspond to high velocity and blue to low velocity. From left to right, a) represents a snapshot of a sample at $t=0$; b) is a snapshot after particle swelling; c) a snapshot after particle erosion. Since we imposed a constant pressure drop, b) has the smallest velocity (hence less erosion) and c) has the highest velocity.}
	\label{fig:porous_example}
\end{figure}

\subsubsection{Circular obstacles}

We consider now circular obstacles on 3D lattice, setting $L_z=1$ and periodic boundary condition along the $z$ direction. Thus, the circles correspond to cylinders in 3D. Along the $x$ and $y$ directions we consider $256 \times 384$ nodes. Along the $x$ direction we impose periodic boundaries and along the $y$ direction we impose a constant pressure drop. The circles are on the central square of the domain, as seen in Fig.~\ref{fig:porous_example}a). The space between the circles and the boundaries ensures that the boundary condition on the $y$ direction, imposed on the top and bottom nodes, does not influence the flow in the pores. The position of the circles was determined using a discrete element simulation where 30 circles start with random position and radius, drawn from a Gaussian distribution, and are compressed by reducing the domain size down to the desired size. We reduce the radius of the compacted circles by half to ensure that there is no overlap after swelling. The nodes overlapped by the compacted circles are converted into solid nodes. The minimum size of the obstacles is half of the initial radius. In total five different configurations of circles with constant initial porosity were simulated. The number of iterations with swelling and erosion depends on the timescale of swelling: $7.5\times\tau_s$.

\begin{figure}
	\centering
	\includegraphics[width=0.5\textwidth]{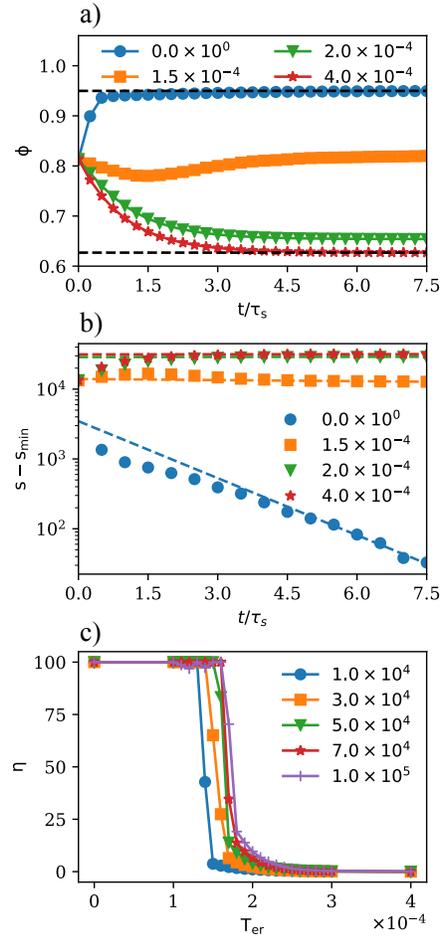}
	\caption{Erosion and swelling of porous media composed of circular obstacles with a constant pressure drop. a) Contains the evolution of the porosity for one of the studied samples, shown in Fig. ~\ref{fig:porous_example}a), for fixed timescale of swelling, $\tau_s=5\times10^4$. The different colors and markers correspond to different erosion thresholds; b) contains the linear fit of the log--linear plot of the solid volume as a function of the reduced time for the parameters used in a); c) contains the average removal efficiency of the five samples studied. The different colors and markers correspond to different timescales of swelling.}
	\label{fig:porous_dp}
\end{figure}

In Fig.~\ref{fig:porous_dp}a), the time evolution of the porosity, for the configuration represented in Fig.~\ref{fig:porous_example} is plotted for different erosion thresholds. Like for the parallel plates we see that, depending on the erosion threshold, the steady state can be characterized by the maximum porosity (blue circles), the minimum porosity (red stars) or an intermediate porosity (green triangles). The orange squares correspond to a simulation that have not reached the steady state. Given the variety of pore sizes, the convergence towards the steady state is slower than the case of parallel plates, especially for cases that are close to total erosion. For these cases, we determine if they fully erode by evaluating how the volume of the solid matrix evolves over the last 10 snapshots (see Fig.~\ref{fig:porous_dp}b)). Given that the volume decreases exponentially fast when erosion dominates, we perform a linear fit of $\log(s(t)-s_\text{min})$ against time. If the linear fit results in a negative slope, then it is assumed total erosion. This results in an efficiency plot, represented in Fig.~\ref{fig:porous_dp}c), with similar behavior to the one of the parallel plates.

\subsection{Constant flow rate}

Imposing a constant flow rate changes how the velocity of the fluid depends on the size of the channel and, consequently, changes the erosion dynamics. For this situation, when the porosity increases, due to erosion, the average velocity decreases which results in lower shear stress. This is reflected on the Hagen--Poiseuille equation for a constant flow rate $Q$
\begin{equation}
	u(r)=\frac{2Q}{{\pi}}\left(\frac{1}{R^2}-\frac{r^2}{R^4}\right) \text{ ,}
	\label{eq:poiseuille_vin}
\end{equation}
and shear stress
\begin{equation}
	T_w=\mu\frac{4Q}{\pi}\frac{1}{R^3} \text{ .}
\end{equation}
When the radius of the tube (porosity) increases the shear stress tends to zero. For the porous media composed of circular obstacles, this implies that when the circles swell it enhances erosion (decreases the pore size), while erosion causes the shear stress to decrease. So, erosion stops when the porosity reaches a threshold value that corresponds to a pore size that is too large for the shear stress to be higher than the erosion threshold. This leads to the conclusion that the steady state does not depend on the timescale of swelling.

\begin{figure}
	\centering
	\includegraphics[width=0.5\textwidth]{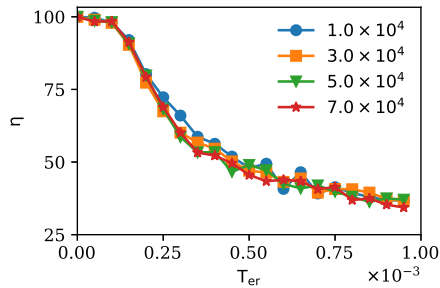}
	\caption{Removal efficiency for the porous medium composed of circular obstacles with a constant flow rate $Q=2.56$. Different colors and markers correspond to different timescales of swelling.}
	\label{fig:porouos_vin}
\end{figure}

We simulated one configuration of circles and fixed the inlet and outlet velocity $\vec{u}_\text{in}=\vec{u}_\text{out}=-10^{-2}\hat{e}_y$ resulting in a fixed flow rate $Q=2.56$. The number of iterations was fixed to be ten times the largest timescale of swelling $7.0\times10^5$. In this way, we ensure that the swelling causes the particle volume to increase by at least $99\%$ and it results in the same time for erosion for all erosion thresholds. Shown in Fig.~\ref{fig:porouos_vin} is the removal efficiency as a function of the erosion threshold. As expected, there is no significant difference in the removal efficiency when the timescale of swelling changes.

\section{Conclusions}
\label{sec:conclusion}

We studied the competition between swelling and erosion in porous media using lattice--Boltzmann simulations. We used Mei's boundary conditions to implement continuous changes in the boundary location and coupled it with an implementation of swelling and erosion. To perform the simulations, we imposed two different boundary conditions at the inlet and outlet: a constant pressure drop and a constant flow rate. Swelling causes an exponential increase of the solid volume, while erosion decreases the solid volume. When the fluid flows due to a constant flow rate, the porosity in the steady state is solely dictated by the erosion parameters. This stems from the fact that erosion causes a decrease in the shear stress on the particle walls, and so eventually the shear stress is smaller than the erosion threshold. When a pressure drop is imposed, the final state is dictated by the competition between swelling and erosion. This is because swelling causes a decrease in the shear stress, while erosion causes an increase. The erosion threshold dictates the dynamics: a small threshold results in an erosion dominated regime, while swelling is the dominant mechanism for large threshold. There is a sharp transition between the two regimes where the porosity, on the steady state, is strongly affected by small changes to the erosion threshold.

Work still needs to be done towards fully understanding the implications of swelling and erosion in porous media. For example, does swelling affect significantly the transport of the eroded mass, and if so, is it reflected on the deposition locations? Other mechanism of interest not addressed in this work was the motion of the particles (bed consolidation) which changes the properties of the flow. The implementation of this mechanism requires the adjustment of the swelling model since one should expect that swelling is affected by the free space in between the particles.

\section*{Acknowledgments}
We acknowledge financial support from the Portuguese Foundation for Science and Technology (FCT) under the contracts no. UIDB/00618/2020, UIDP/00618/2020, and SFRH/BD/143955/2019.

JSA acknowledges the Brazilian agencies CNPq, CAPES, and FUNCAP for financial support.

\bibliography{dsfd_ref}

\end{document}